\documentstyle[epsf,aps,fleqn]{revtex}

\newenvironment{figures}[1]%
{\begin{list}{}{\settowidth{\labelwidth}{#1}
  \setlength{\leftmargin}{\labelwidth}
  \addtolength{\leftmargin}{\labelsep}
  \setlength{\parsep}{1ex plus0.7ex minus0.7ex}
  \setlength{\itemsep}{0.8ex}
  }}{\end{list}}

\def\mc#1 {\multicolumn{1}{|c|}{#1}}

\newcommand{\beq}{\begin{equation}}
\newcommand{\eeq}{\end{equation}}
\newcommand{\beqa}{\begin{eqnarray}}
\newcommand{\eeqa}{\end{eqnarray}}

\begin{document}
\vspace{1cm}
\title {
\hspace{11cm} hep-ph/9610528 \\ \vspace{1cm}
{\LARGE {\bf Monte Carlo calculations for the hard Pomeron}}} 

\author{
L.P.A. Haakman
$^{a}$,
O.V. Kancheli
$^{b}$,
J.H. Koch
$^{a,c}$}

\address{
$^{a}$
National Institute for Nuclear Physics
and High Energy Physics (NIKHEF), \\
P.O. Box 41882, NL-1009 DB Amsterdam, The Netherlands \\
$^{b}$
Institute of Theoretical and Experimental Physics, \\
B. Cheremushinskaya 25, 117 259 Moscow, Russia  \\
$^{c}$
Institute for Theoretical Physics,\\
University of Amsterdam, Amsterdam, The Netherlands\\
}


\maketitle

\begin{abstract}
Starting from the same input as the standard
BFKL Pomeron, we directly calculate the ``hard'' Pomeron
as a gluonic ladder by using Monte Carlo methods.
We reproduce the characteristic features of the the BFKL 
Pomeron and are now also able to evaluate new observables.
The applicability of the BFKL approach under realistic kinematical
conditions can be tested and the influence of the running coupling
constant examined.
\end{abstract}

\section{Introduction}
A large amount of data for reactions at high energies could be
explained very succesfully in terms of the exchange of the
Pomeron.
It was introduced into the general framework of reggeon exchanges
in an {\it ad hoc} fashion and provides the dominant contribution.
It is therefore a major challenge to understand the nature of this Pomeron
microscopically in terms of QCD.
Most of the data so far involved small momentum transfers, and the
exchanged Pomeron was referred to as a ``soft'' Pomeron.
In trying to understand this phenomenology on the basis of QCD, it
can be argued that the underlying mechanism mainly consists of the
exchange of gluons, is of long range and thus involves
non-perturbative dynamics.

Attempts to understand the Pomeron from a perturbative QCD point of
view lead to the concept of a ``hard'' Pomeron
that is involved in small distance, high momentum transfer processes.
The interest in this Pomeron was renewed when recent data from HERA
at low $x$ and large $Q^2$ showed features that could be explained
with the exchange of such a hard Pomeron, which has a behavior quite
different from the soft one.
The theoretical approach to understand the hard Pomeron goes back to the
work of Lipatov {\it et al.} (BFKL) \cite{BFKL,Lipatov}.
These authors were able to infer some properties of the hard Pomeron perturbatively, describing it in terms of a reggeized gluon ladder.
Their calculation was performed in leading logarithmic order
$(\alpha_s \log s \sim 1$ for a fixed $\alpha_s \ll 1$ ) where the
multiregge kinematics gives the dominant contribution.

In this paper, we use a numerical method to better understand the 
features of the hard Pomeron and to see to what extent the concept of a 
perturbative Pomeron as embodied in the BFKL approach remains
valid.
Rather than proceeding as BFKL and solving analytically under 
the conditions mentioned above, where $\log s$ goes 
to infinity, we use a Monte Carlo method to directly evaluate
reggeized gluon ladders over a broad kinematical range and for
different values of $\alpha_s$.
This allows us to study the ladder structure explicitly and in 
more detail.
In particular, we are interested to see to what extent the kinematics 
within the ladder stays in a region where perturbative QCD can be
applied.
Our method also allows us to examine the consequences of letting the
coupling constant run.
Our approach differs from previous numerical studies which use angular
ordering \cite{Marchesini} or start from the BFKL equations \cite{Bartels}.

\section{The BFKL Pomeron}
We will shortly review the approach leading to the BFKL Pomeron.
The BFKL Pomeron can be viewed as the exchange of
gluonic ladders between two particles $A$ and $B$ as shown in Fig.1.
The exchanged ladder must be a color singlet state.
Working in the leading logarithmic approximation, the main
contribution then comes from the multiregge region.
In terms of the rapidities $y_i$ of the emitted gluons with momenta
$k_{i}$, in this kinematical region the rapidity intervals, $\delta y_i$, between two neighboring emitted gluons with rapidities
$y_{i-1}$ and $y_{i}$, satisfy $ \delta y_{i} \gg 1$.
The BFKL Pomeron thus is a gluon ladder where the rapidities along
the ladder increase monotonically.
The total rapidity interval, $Y = Y_B - Y_A$, between the
particles $A$ and $B$ can be simply expressed as the sum
of the individual rapidity intervals:
$Y = \sum_{i=1}^{n+1} \delta y_i~~.$

The main features of the BFKL Pomeron can be discussed by starting
from the expression for the total inclusive cross section for the
reaction $A + B \rightarrow X$, proportional to the imaginary part
of the elastic amplitude with a gluon ladder exchange; an
example with $n$ rungs is shown in Fig.1.
The differential cross section has the form
\beq
d\sigma_{n}(A+B \rightarrow X) =
\frac{\pi}{s^2}\prod_{i=1}^{n+1}
\frac{dy_i~d^2{\bf k}_{i \perp}}{2(2\pi)^3}
\delta\left(Y-\sum_{i=1}^{n+1} \delta y_i\right) |T_n|^{2}~,
\label{ImAmpli}
\eeq
where $T_n$ is the amplitude for emission of $n$ hard gluons (gluon jets).
The Pomeron exchange consists of the sum over all such ladders with 
different numbers of gluons.
The main ingredients in this gluon emission amplitude are the non-local,
gauge invariant effective gluon vertices, $\Gamma^{\mu}$,
and the reggeized gluon propagators.
In the cross section, the vertices appear only in
a form contracted over color indices and gluon polarizations,
{\it i.e.} as a contraction of two ``Lipatov vertices'',
\beq
\Gamma^{\mu}({q}_k,{q}_{k+1}) \cdot
\Gamma_{\mu}({q}_k,{q}_{k+1}) = 4
\frac{({\bf q}_{k\perp}^2+\mu^2)({\bf q}_{k+1\perp}^2+\mu^2)} {({\bf
q}_{k\perp}-{\bf q}_{k+1\perp})^2+\mu^2}~~.
\label{LipVertex}
\eeq
The gluon mass ${\mu}$ is introduced
in order to regulate the infrared behaviour.
The Lipatov vertices can only be used in the case of 
large rapidity intervals because otherwise higher-order corrections
to the vertices become important.
The reggeization of the gluons is taken into account by replacing
the free gluon propagators by the reggeized ones:
\beq
\frac{1}{t_{i}-\mu^2}~~\longrightarrow~~
\frac{{\rm e}^{\delta y_{i}\epsilon(t_{i})}}{t_{i}-\mu^2}
\label{Reg.prop.}~~,
\eeq
where $t_i = q_i^2 \simeq -{\bf q}^2_{i_\perp}$ and $\epsilon$ the
gluon trajectory.
For large momenta $\epsilon$ is given by
\beq
\epsilon({\bf q}_\perp^2) \sim -\frac{\alpha_s N_c}{4\pi} \log\left(
\frac{ {\bf q}_\perp  ^2}{\mu^2}\right)~~.
\label{Traj}
\eeq
The reggeized gluons represent the exchange of color octet
gluon ladders.
Their use can be interpreted as taking into account Sudakov-like
suppression; the inclusion of radiative corrections corresponds
to the probability that there is no emission of additional gluons
in rapidity interval $\delta y_i$.
In this way any double counting is avoided.
Since the infrared divergences coming from the vertices and the 
regggeized propagators cancel each other, one can safely take
the limit $\mu\rightarrow 0$ as is done in the original BFKL Pomeron.
It can be shown that the colour octet part of gluon ladder
exchanges as in Fig.1 indeed yields the reggeized gluon, proving the
self-consistency of this approach.

The reaction mechanism that is independent of the properties of the
inital particles $A$ and $B$, the exchanged Pomeron, is indicated by the
dashed box in Fig.1.
For the dashed box one can derive a recursion relation in the number
of the emitted gluons  which
leads to the well-known BFKL equation \cite{BFKL,Lipatov}.
For the cross section for this subprocess, the interaction
of two gluons (with transverse momenta ${\bf q}_{A\perp}$ and
${\bf q}_{B\perp}$),
the BFKL equation predicts in the asymptotic limit
\beq
\frac{d\sigma}{d{{\bf q}^2_{A \perp}} d{{\bf q}^2_{B\perp}}} \sim
\frac{1}{\sqrt Y} \exp\left(Y\Delta - \frac{\log^2({\bf q}_{A
\perp}^2/{\bf q}_{B \perp}^2)}{4 B Y}\right)~,
\label{diff.crs}
\eeq
where
\beq
\Delta={ \alpha_s \frac{4 N_c}{\pi}
\log 2} \simeq 2.65 \cdot \alpha_s ~~,~~ B=14~\zeta(3)
~\frac{\alpha_s N_c}{\pi} \simeq 12.6 \cdot \alpha_s ~.
\label{DeltaB}
\eeq
It is necessary that the initial and final transverse momenta
of the ladder are large such that perturbation theory can 
be expected to hold.
This means that the dimensions of the hadrons $A$ and $B$
($R_A$ and $R_B$) should be small.
From Eq.(\ref{diff.crs}) one can now see several characteristic
features of the BFKL Pomeron exchange:
\begin{itemize}
\item{} {\it Energy dependence of the total cross section:}\\
After integration over the gluon transverse momenta, one obtains
for $\sigma$
\beq
\sigma \sim \frac{1}{\sqrt Y} \exp({\Delta Y}) \sim s^{\Delta}~.
\label{tot.crs}
\end{equation}
For a realistic value of the fixed coupling constant, $\alpha_s=0.2$,
one obtains for $\Delta=0.53$, which is much larger than for the
soft Pomeron, while the low-$x$ data indicate $\Delta\simeq 0.3$.
So it will be interesting to find out what will be the effect of
including a running coupling constant in the BFKL Pomeron
on the intercept.
\item {} {\it Diffusion pattern of the transverse momentum
distribution:}\\
From Eq.(\ref{diff.crs}) one also immediately
recognizes the diffusion behavior in the logarithm of the
transverse momenta {\cite{Bartels}}.
For a gluon with transverse momentum ${\bf q}_{\perp}$ and with 
rapidity $Y_A+y$ the logarithm of the transverse momentum will 
under the condition $1 \ll y \ll Y$ fluctuate around its initial value 
$\log({\bf q}_{A \perp}^2)$ according to
\begin{equation}
<|\log({\bf q}_\perp^2)-\log({\bf q}_{A\perp}^2)|>^2_y 
\simeq C {y} ~~~~
,~~~~ C = \frac{4B}{\pi} \simeq 16.0~\alpha_s~~.
\label{logkt}
\end{equation}  
The transverse momenta are not ordered like in the Altarelli-Parisi 
evolution, but they perform a random walk.
This means that there can be contributions to the ladder from
configurations both with low and high transverse momenta. 
For the relatively small starting ${\bf q}_{A \perp}^2$ we 
consider below, the momenta diffuse quite rapidly to larger values
of ${\bf q}_{\perp}^2$.
\item {} {\it Rapidity distribution:}\\
Eq.(\ref{diff.crs}) can also be used to obtain the gluon density
in the ladder at a given rapidity difference $y$.
In Refs.\cite{DelDuca93,Ryskin} the inclusive cross section for
the production of a gluon (gluon jet) with transverse momentum
${\bf k}_{\perp}$ and rapidity $Y_A + y$ in the exchange of a 
BFKL Pomeron was derived for the case  $Y - y \gg1$ and $y \gg 1$.
Under these conditions the emitted gluon is not too close to the 
ends of the ladder.
This cross section corresponds to fixing the momentum of a rung in the gluon ladder of Fig.1 and can be described by Eq.(\ref{diff.crs}) for the parts
of the ladder above and below the gluon in question.
Upon integration over ${\bf k}_{\perp}$, one can obtain the rapidity
distribution $\rho (y)$, the gluon density per unit rapidity interval of 
the emitted gluons at a given rapidity difference $y$,
\beqa
\rho(y)&&\equiv\int {d{\bf k}_\perp^2}~\frac{1}{\sigma}\frac{d\sigma}
{dy d{\bf k}_\perp^2}\left[g g \rightarrow X g(y,{\bf k}_\perp^2)\right]
{\nonumber}\\ 
&&\simeq 1.6~ \alpha_s^{3/2} \sqrt{\frac{y(Y-y)}{Y}}~~.
\label{inclusive}
\eeqa
In the center of the spectrum one has
$\rho(y=Y/2)\simeq 0.8 \cdot \alpha_s^{3/2} \cdot \sqrt{Y}$.
The BFKL results are valid only in multiregge kinematics,
where  $<\delta y>_y=\rho^{-1} \gg 1$.  
To fulfill this condition one should have
$\alpha_s Y \ll 1.2 \alpha_s^{-2}$.
At the edge of the inclusive spectrum ($y \ll Y$) one finds
\beq
\rho(y) \simeq
1.6~\alpha_s^{3/2}~\sqrt{y} + D(\alpha_s)
~~,\label{gluondensity}
\eeq
where an unknown $y$ independent function $D$ 
which determines the density at small $y$, has been added.
The rapidity interval between neighboring gluons $<\delta y>_y
\simeq 0.6/(\alpha_s^{3/2}\sqrt{y}) $~~ becomes thus
smaller with $y$ towards the middle of the ladder.
Note that the analytical result in Eq.(\ref{gluondensity}) can be 
simply understood. 
As the only dependence of the amplitude $T_n$ on $\delta y$ is
due to the propagator, Eq.(\ref{Reg.prop.}), this $\rho(y)$
can be estimated as
$\rho (y)\simeq 1/<\delta y>_y\sim 2\epsilon(<{\bf q}^2_\perp>)\sim
\alpha_s^{3/2} \sqrt{ y}$,
where we used Eq.(\ref{logkt}) to approximate
$<\log{\bf q}^2_\perp>\sim \sqrt{\alpha_s y}$.

\end{itemize}

\section{Monte Carlo approach for the Pomeron}
To get a more detailed understanding of the hard Pomeron,
a direct evaluation of the multidimensional integrals in 
Eq.(\ref{ImAmpli}) would be necessary.
The expected gluon multiplicities for large total rapidity intervals
are large, $n \sim 10 \div 100$, and thus the integrals of such 
high dimensionality that straightforward evaluation is impossible.
We therefore have developed a Monte Carlo approach which enables
us to deal explicitly with the multidimensional integrals.
By keeping track of the kinematics, we can then examine to what
extent the BFKL Pomeron indeed is made up from hard gluons and, for
example, if the assumptions about the large rapidity intervals,
$\delta y_i \gg  1$, are justified throughout the ladder.
An advantage of working numerically is that we can investigate
what effect the use of a running coupling constant will have 
on the structure of the gluon ladder.
We use for the running coupling constant 
\beq
\alpha_s({\bf q}_\perp^2)=\frac{1}{b_0\log\left
(a+{\bf q}_\perp^2/\Lambda^2\right)}~~, \label{alfas}
\eeq
{\it i.e.} the standard expression with $\Lambda=250$ MeV, but
extended by an additional parameter $a$ that fixes the 
coupling constant at $|{\bf q}_\perp|=0$, thus regularizing
the infrared singularity.
The parameter $a$ can be interpreted as the interaction with 
non-perturbative vacuum fluctuations which freezes the coupling
constant at low momentum scales \cite{Simonov,Hancock}.
The value of $a$ should be such that $\alpha_s (0)\sim 0.4 \div 0.7$.
In our calculations we took $a=10.3$, corresponding to $\alpha_s(0)=0.6$~   .

As in the discussion in the last section, we will focus on the
gluon subprocess indicated by the box in Fig.1.
Therefore, we will generate for our Monte Carlo evaluation
configurations $\{n, y_i,{\bf k}_{i\perp}\}$ for the emitted gluons or,
equivalently, for the intermediate gluons along the ladder, separated
in rapidity by at least $\delta y_{min} \geq 0$, which is an input parameter.
This is done in two steps.
First, we generate ladders with a certain distribution in the momenta and
in the number of rungs by an approximation method as described below.
Each of these configurations is then used as the starting point
for a Metropolis procedure to generate new configurations.
This serves to correct for errors inherent in the approximation 
in the first step. 
This procedure leaves the distribution in $n$ unchanged.

In generating these multidimensional configurations, we make use of
the specific dependence of the integrand on the kinematical variables,
\beqa
&&|T_{n}|^{2}~
\sim \prod_{i=1}^{n}f(\delta y_{i+1},{\bf q}_{i\perp},{\bf
q}_{i+1\perp}) ~~,~ \label{FactAmp} \\
&&f(\delta y_{i+1},{\bf q}_{ i \perp},{\bf q}_{i+1
\perp})= \frac{3\alpha_s({\bf q}_{i+1 \perp}^2)}{\pi^2} \frac{{\rm
 e}^{2\delta y_{i+1}\epsilon ({\bf q}_{i+1 \perp}^2)}}{({\bf q}_{i
 \perp}-{\bf q}_{i+1 \perp})^2 +\mu^2}~~. \label{kernel}
\eeqa
Note that the external kinematics requires ${\bf q}_{1 \perp} =
{\bf q}_{A \perp} $ and ${\bf q}_{n+1 \perp} ={\bf q}_{B \perp}$.
Further we take for the gluon trajectory the usual form
\beq
\epsilon({\bf q}_\perp^2)=- ({\bf q}_\perp^2+\mu^2)
\int \frac{d^2 {\bf
k}_\perp}{(2\pi)^2}\frac{N_c~ \alpha({\bf k}_\perp^2)}
{({\bf k}_\perp^2+\mu^2)([{\bf k}_\perp-{\bf q}_\perp]^2 +\mu^2)}~~, \label{trajectory}
\eeq
but now with a running coupling constant and which at high ${\bf q}_\perp^2$
for fixed coupling constant leads to the asymptotic form of Eq.~(\ref{Traj}).

The fact that Eq.(\ref{FactAmp}) is a product of functions of pairs
of kinematical variables referring to subsequent rungs in the ladder
allows one to generate the ensemble of configurations in a sequential
fashion, for which we chose the von Neumann rejection method.
It is essential that the factors $f(\delta y_{i+1},{\bf q}_{ i \perp}
,{\bf q}_{i+1\perp})$ are positive definite and thus can be
interpreted as probabilities for the production of the
$({i+1}){\rm th}$ rung with rapidity difference
$\delta y_{i+1}$ and linked to the former rung by a gluon 
with momentum ${\bf q}_{i+1\perp}$, when the previous rungs
have already been generated.
In fact, the conditional probability to produce the
$(i+1){\rm th}$ rung with
$(\delta y_{i+1},{\bf q}_{i+1\perp})$ as part of the whole ladder
is given by
\beq
f(\delta y_{i+1},{\bf q}_{i \perp},{\bf q}_{i+1 \perp})
\cdot \psi(Y-Y_{i+1},{\bf q}_{i+1 \perp})~~,
\label{cond.prob}
\eeq
where $Y_{i+1} = \sum_{k=1}^{i+1}\delta y_k$ is the rapidity interval
between the `zeroth' ({\it i.e.} particle A)
and $(i+1){\rm th}$ rung and
\beq
\psi(Y-Y_{i+1},{\bf q}_{ i+1 \perp}) =
\sum_{n=i+1}^{\infty}\int\prod_{l=i+2}^{n+1} dy_{l}~d^2 {\bf q}_{l
\perp} f(\delta y_{l},{\bf q}_{l-1 \perp},{\bf q}_{l
\perp}) \delta\left(Y-Y_{i+1}-\sum_{l=i+2}^{n+1} \delta y_l\right)
\label{psi}
\eeq
is the total probability of configurations `after' this
$(i+1){\rm th}$ rung.
Since the integrand in $\psi$ is given by a similar factorized form
as Eq.(\ref{FactAmp}), we can use Eq.(\ref{diff.crs}) to see
that at very large $Y-Y_{i+1}$
\beq
\psi(Y-Y_{i+1},{\bf q}_{ i+1 \perp}) \sim  \exp{\left(-\frac{\log^2(
{\bf q}_{i+1\perp}^2/{\bf q}_{B\perp}^2)}{4B(Y-Y_{i+1})}\right)}~~,
\eeq
where ${\bf q}_{B\perp}$ is the transverse momentum of the final
gluon in the ladder which is taken to be fixed and not too large.
So for $Y-Y_{i+1} \gg Y_{i+1}$~~ the function ~$\psi$~ is constant
for  values of ~$\log {\bf q}^2_{i+1} \sim \sqrt{Y_{i+1}}$~
which we expect to be essential in the generation of
the quantities $(\delta y_{i+1},{\bf q} _{i+1})$ at the
$(i+1){\rm th}$ rung.
Therefore when generating the initial part of the ladder, it is a good approximation  to use only the first factor
$f(\delta y_{i+1},{\bf q}_{ i \perp},{\bf q}_{i+1\perp})$
in Eq.(\ref{cond.prob}).
Starting with the first rung, for which we take $|{\bf q}_{A \perp}| = R_A^{-1}$, we can therefore generate subsequently for every
rung the variables $(\delta y_{i},{\bf q}_{i \perp})$.
We repeat this until $Y_{n+1} \simeq Y$ , which in our approach 
then fixes the gluon multiplicity $n$ in this ladder.
We stress that we will only use the first part of the ladder
for the results below.

It should be noted that when using the weight functions generated with
the von Neumann rejection method, we have to introduce an ultraviolet cut in ${\bf k}_{\perp}$.
This is due to neglecting $\psi$ in Eq.(\ref{cond.prob}).
The dependence on this cutoff parameter $Q_{max}$ is of type
$\log\log (Q_{max}^2/\mu^2)$ and thus weak.
Since we fix the end momentum of the ladder, $|{\bf q}_{B\perp}|\sim
R_{B}^{-1}$, the dependence on $Q_{max}$ vanishes.

For the Metropolis routine, we use the configurations obtained
by the von Neumann method as starting configurations to
improve the sampling in the phase space, producing
a series of new configurations $\{n,y'_i,{\bf k'}_{i \perp}\}$.
Note that the number of rungs $n$ is not changed.
A more detailed discussion  will be published elsewhere.

To calculate transverse momenta distributions,
mean rapidity intervals {\it etc.}, we take a very large
total rapidity interval $Y$ and consider only a part of the
ladder for which ${y} \ll Y$ in accordance to our approximation
$\psi \sim constant$.
That this yields predictions independent of $Y$ and other quantities
characterizing the particle $B$ at the end of the gluon chain
can also be seen, for example, from the asymptotic expression for the
inclusive cross section, Eq.(\ref{inclusive}).
So one should realize that using our method we can investigate
only distributions in the growing part of the gluon spectrum.

For the dependence of the total cross section on the
energy $s$ we have to proceed differently.
In order to obtain the Pomeron intercept it is sufficient for the
total cross section to integrate (\ref{FactAmp}) over the full
phase space with logarithmic precision.
This was done in two ways.
In the first method, we used the ratio of accepted
to total events in the von Neumann procedure at different values of 
$Y \sim \log s$.
Fitting this to an exponential function, we obtain the exponent $\Delta$ 
in Eq.(\ref{tot.crs}).
In the second method, we first calculate for a given energy $s$
(or rapidity interval $Y$) the mean multiplicity of the ladder and 
then approximate the Pomeron by a ladder containing that many rungs.
In the corresponding amplitude, Eq.(\ref{FactAmp}), we then replace 
the transverse momenta of the previous rungs in the products of 
the functions $f(\delta y_{i+1},{\bf q}_{ i \perp},{\bf q}_{i+1\perp})$
by the already determined mean value for that rung.
This approximation enables us to compute all integrals needed for 
the total cross section separately.
Performing this again for different values of $Y$ the exponent $\Delta$
can be extracted.
Both methods were found to agree reasonably well.

\section{Results and Discussion}

We first show what our Monte Carlo calculations yield for the
properties of the gluon ladder discussed in Section $2$
and compare them to the predictions of the BFKL theory in the
asymptotic limit.
In all our calculations presented in this section we chose
for the ``constituent'' gluon mass $\mu$ one half of the 
gluonium mass: $\mu \simeq 0.5$ GeV.
We have checked that changes by a factor of two in this parameter
had a negligible effect on the results below. 
Furthermore, we established in all cases that there is indeed 
no dependence on the ultraviolet cutoff introduced in the von Neumann step.

First we examined the exponent $\Delta$, which fixes
the Pomeron intercept.
We performed calculations with a fixed coupling constant
in the range $0.01$--$0.20$ and for a total rapidity interval 
$Y=50$.
For the lowest values of $\alpha_s$ this yields rather short ladders
($n \sim 6$) for which our method of generating rungs in the ladder
is less reliable.
For $\alpha_s = 0.05$, we deal with ladders of total length $n\sim 2 0$ and
obtain $\Delta = 0.11$, close to the BFKL result of Eq.(\ref{tot.crs})
which gives $0.13$.
Taking $\alpha_s = 0.1$, we get numerically $0.27$, compared to the BFKL
value of $0.27$.
This reasonable agreement indicates that our numerical approach is 
quite reliable.

To use our samples to extract the gluon transverse momentum and 
rapidity distributions, we should restrict ourselves to the first
part of the ladder, {\it i.e.} up to rapidity intervals of ${y}\sim 20$ 
in our example.
For the final gluon we took $|{\bf q}_{B\perp}| = R_B^{-1}
\simeq R_A^{-1}$.
In this special case with similar transverse momenta for both ends 
the last part of the ladder is identical to the first part and,
 aside from some uncertainty in the middle of the ladder, the complete ladderstructure can be obtained. 
Results for the mean logarithm of the transverse momenta for an initial gluon
momentum of $|{\bf q}_{A\perp}| = R_A^{-1} = 10 ~{\rm fm}^{-1}$ and 
for two values of the coupling constant $\alpha_s$ are shown in
Fig.2. 
Indeed we see that $<|\log({\bf q}_\perp^2/{\bf q}_{A\perp}^2)|>^2$
follows the behavior predicted by the BFKL theory, Eq.(\ref{logkt}):
a linear rise with the rapidity difference $y$.
The slopes also agree reasonably with the BFKL prediction. 
For $\alpha_s=0.05$ and $\alpha_s=0.1$ we find respectively $0.31$ and
$0.51$ compared to the analytic BFKL values of $0.8$ and $1.6$.
The gluon density $\rho(y)$ is shown in Fig.3.
Here our results also reproduce the
BFKL predictions well, both the $y$ dependence and the
coefficients; we fitted both curves to the function $\rho(y)= C_0 +C_1
\sqrt{y}$ and found for $C_1$ a value of $2.2~10^{-2}$ for 
$\alpha_s=0.05$ and $5.8~10^{-2}$ for  $\alpha_s=0.1$ wheras 
BFKL in the asymptotic limit predicts $1.8~10^{-2}$ and $5.1~10^{-2}$, respectively.

This good agreement with the main BFKL features confirms
the validity of our method.
We now use it to examine several aspects that go beyond
the original framework of the BFKL Pomeron.
The first concerns the effect of a running coupling constant.
We repeated the calculation for the total cross section with the
$\alpha_s$ given in Eq.(\ref{alfas}).
If we try to fit the energy dependence according to Eq.(\ref{tot.crs}),
the effective value for $\Delta$ clearly decreases compared to that
obtained with a fixed coupling constant $\alpha_s=\alpha_s({\bf q}_{A\perp }^2)$, but our method of calculating
is not reliable enough to give its precise value .
This decrease can be understood from Fig.2, where we also show a curve for
a running coupling constant.
We observe that ${\bf q}_\perp^2$ continues to grow as we go down
the ladder (just as for fixed $\alpha_s$);
this is due to our choice of ${\bf q}_{A\perp}^2$
which leads to a rapid diffusion from $\log({\bf q}_{A\perp}^2)$
mainly to larger values of $\log({\bf q}_{\perp}^2)$.
This leads to contributions with lower coupling constants, resulting in
a lower $\Delta$. 
The effect of a running coupling constant on the gluon density
is shown in Fig.3. While there are some changes, it has
no essential influence on the qualitative features of the
distribution of gluons in the hard Pomeron.
There have been previous studies of the effect of a running
coupling constant \cite{Lipatov,Hancock,Kwiecinski}. 
They showed the decrease of $\Delta$ and its influence on the character
of the Pomeron singularity.

Another important aspect we can investigate is the rapidity 
interval, $\delta y$, between successive rungs of the gluon ladder.
What we show in Fig.4 is the average rapidity interval as
function of the rung number in the ladder.
Our results show a rapid drop of $<\delta y>$ for the first
rungs, going over in a much slower decrease which continues to
the end of the ladder.
In our calculations we imposed a lower bound of $\delta y_{min} = 0$,
since the leading contribution is assumed to come from high 
$\delta y$ configurations.
Fig.4 makes clear how important the value of this cut is:
a major part of the ladder consists of rungs with only
a relatively small rapidity separation, typically of order $1.5$ for
$\alpha_s = 0.1$.
Thus had we chosen a large $\delta y_{min}$, entirely different 
results would have been obtained.
The use of a running coupling constant leads to the same
observations.
For a smaller fixed coupling constant ({\it e.g. }$\alpha_s\sim 0.05$)
we find rapidity intervals mainly around $3.0$, which means
imposing $\delta y_{min} > 0$ would have a smaller effect
on the ladder structure.
From this analysis we see that configurations with small
rapidity intervals in the ladder play an important role and can influence
the results for larger values of the fixed coupling constant and for 
the running one.


Our Monte Carlo study of the hard Pomeron in terms of explicit 
calculations of gluon ladders reproduced the main predictions 
of the BFKL Pomeron over a wider range of kinematical conditions.
This seems to confirm the validity of the kinematical assumptions 
that go into the BFKL derivation.
However, when one looks in more detail at the gluon ladder, which 
can be done with our numerical method, it becomes clear that the
multiregge kinematics are not applicable anymore towards the middle 
of the ladder, especially for not too small values of $\alpha_s$.
In a forthcoming publication, we will therefore study the validity
of the BFKL Pomeron in more detail for different kinematical
conditions and the importance of higher order corrections through
numerical simulations.
Theoretical progress has been made in computing the next-to-leading
order terms \cite{Fadin}, but the expressions are rather complicated 
and there is presently little hope that higher order terms can be computed
exactly.
More intuitive pictures for the hard Pomeron such as our direct ladder
calculations or the color dipole model \cite{Mueller} can show the 
way towards a better understanding of the nature of the ``hard'' Pomeron.

\vspace*{1cm}
{\bf  ACKNOWLEDGEMENTS}

We thank K. Boreskov and A. Kaidalov for stimulating discussions,
and Y. Simonov and K. Ter Martirosyan for interesting remarks.
The work of L.H. and J.K. is part of the research program
of the Foundation for Fundamental Research of Matter (FOM)
and the National Organization for Scientific Research (NWO).
The collaboration with between NIKHEF
and ITEP was supported by a grant from NWO and by grant 93-79 of
INTAS.
O.K. also acknowledges support from grant J74100 of
the International Science Foundation and the Russian Government.

\small{
 }
\newpage

\section*{\bf Figure Captions}

\begin{figures}{}

\item[{\bf Fig.1}] The contribution of a gluonic ladder with $n$ rungs
to the Pomeron exchange.
The Lipatov vertices are denoted by the dots and the reggeized
gluon propagators by the thick, vertical gluon lines.
\item[{\bf Fig.2}] The square of the mean for the absolute value 
of the logarithm of the gluon transverse momentum along the ladder
, $|\log({\bf q}_\perp^2/{\bf q}_{A\perp}^2)|$, 
as function of rapidity interval ${y}$ for different
coupling constants;
the dashed line for $\alpha_s=0.05$, the solid line for $\alpha_s=0.1$
and the dot-dashed line for a running coupling constant.
\item[{\bf Fig.3}] The gluon density, $\rho$, as a function of rapidity 
interval ${y}$ for different coupling constants;
the dashed line for $\alpha_s=0.05$, the solid line for $\alpha_s=0.1$
and the dot-dashed line for a running coupling constant.
\item[{\bf Fig.4}] The average rapidity interval between two
neigboring rungs as function of the rung number $N$ for different 
coupling constants;
the dashed line for $\alpha_s=0.05$, the solid line for $\alpha_s=0.1$
and the dot-dashed line for a running coupling constant.
\end{figures}
\newpage
\vspace*{4cm}
\centerline{\epsfysize=15cm \epsffile{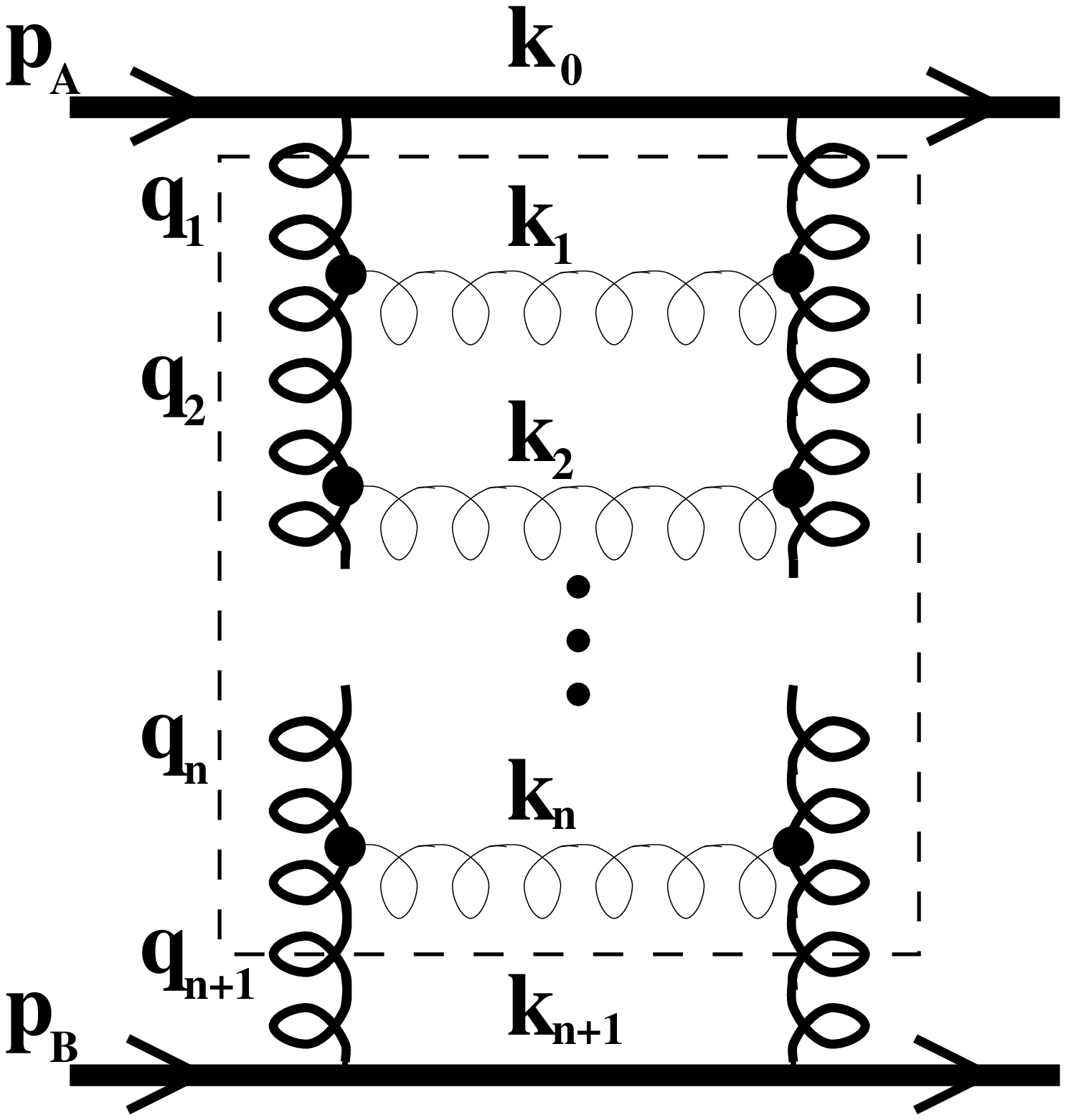}}
\nopagebreak
\begin{center}
{\LARGE {\bf Figure 1} }
\end{center} \vspace*{0.5cm}
\newpage
\vspace*{4cm}
\centerline{\epsfysize=15cm \epsffile{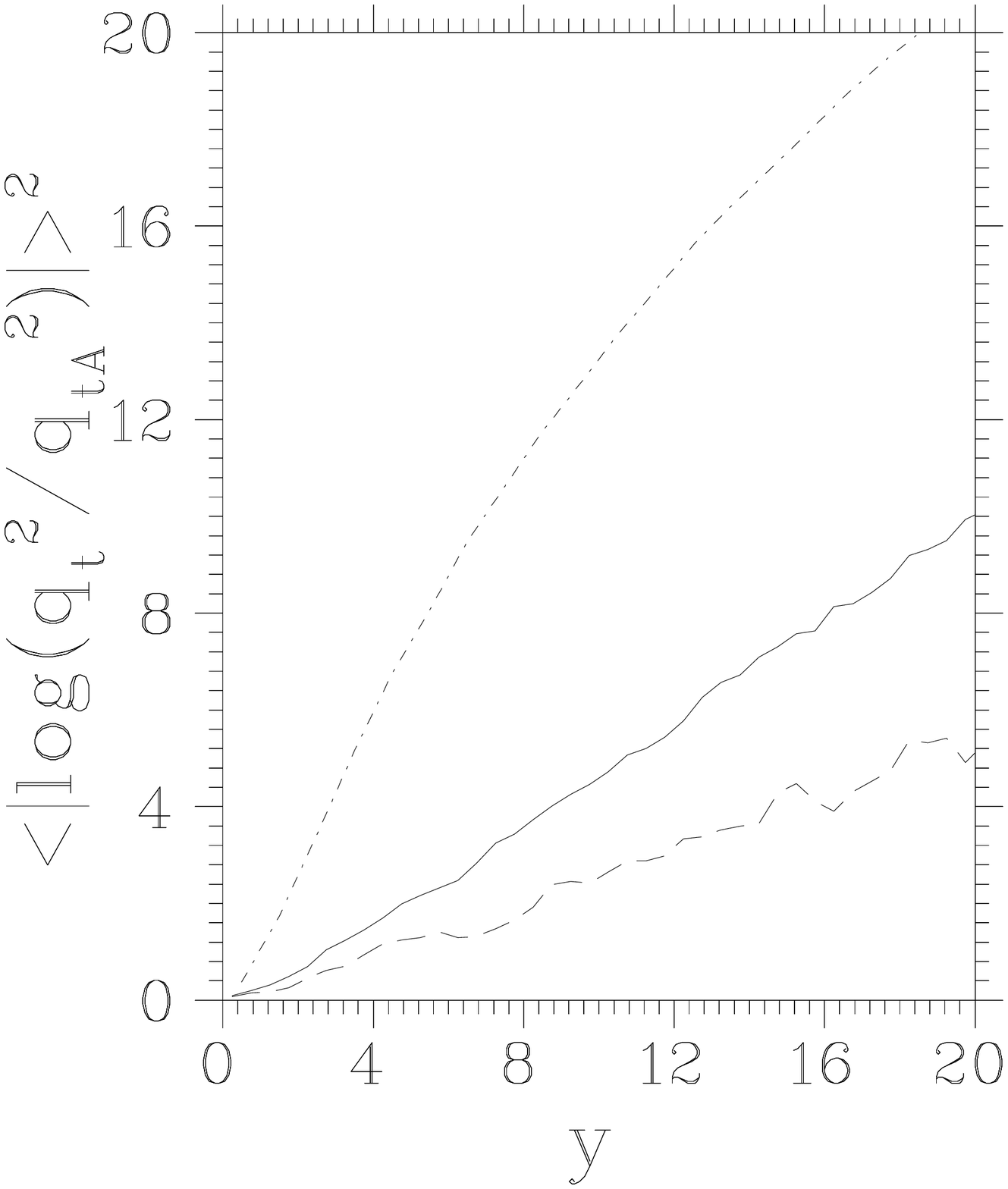}}\nopagebreak
\begin{center}
{\LARGE {\bf Figure 2} }
\end{center} \vspace*{0.5cm}
\newpage
\vspace*{4cm}
\centerline{\epsfysize=15cm \epsffile{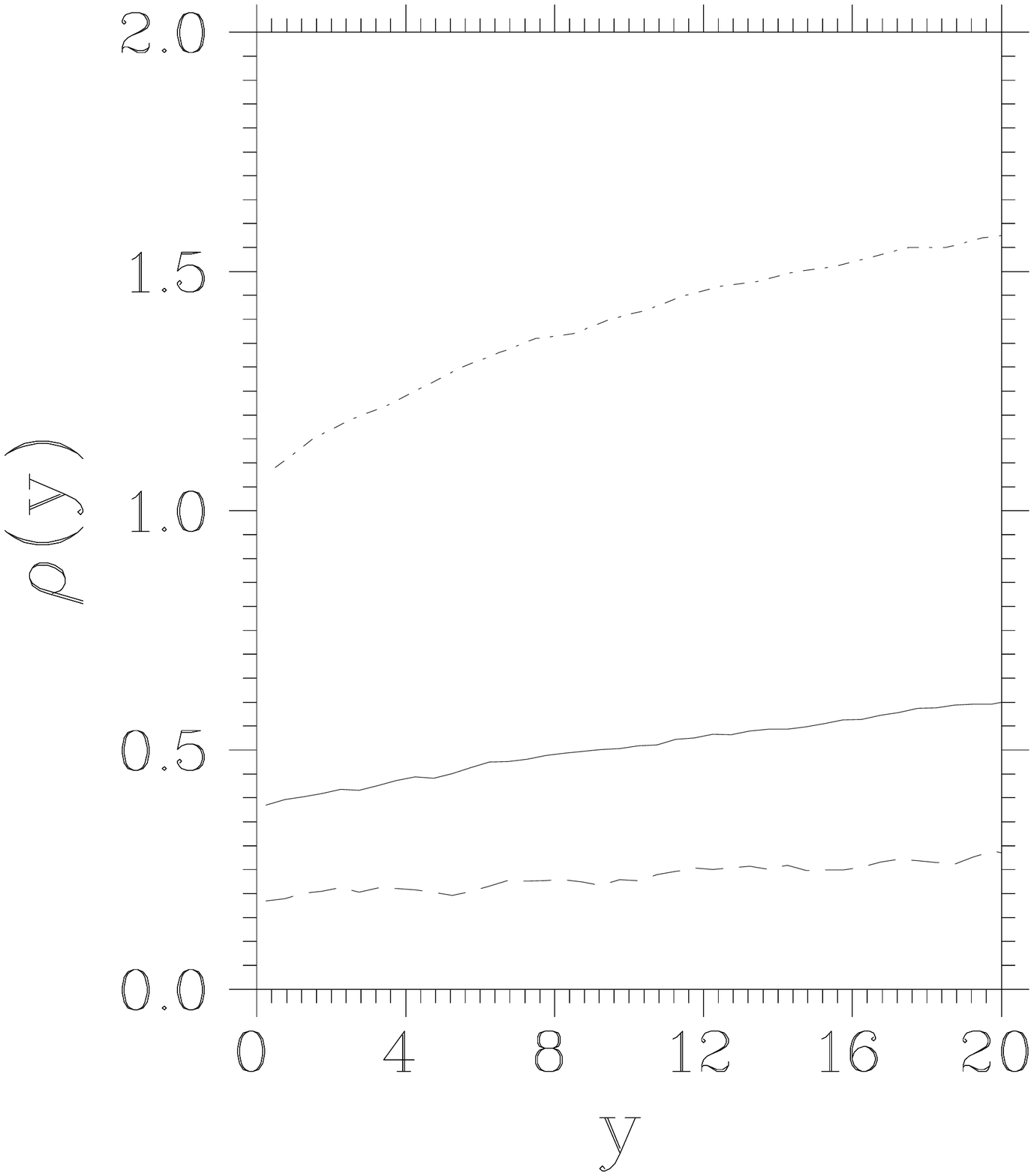}}
\nopagebreak
\begin{center}
{\LARGE {\bf Figure 3} }
\end{center} \vspace*{0.5cm}
\newpage
\vspace*{4cm}
\centerline{\epsfysize=15 cm \epsffile{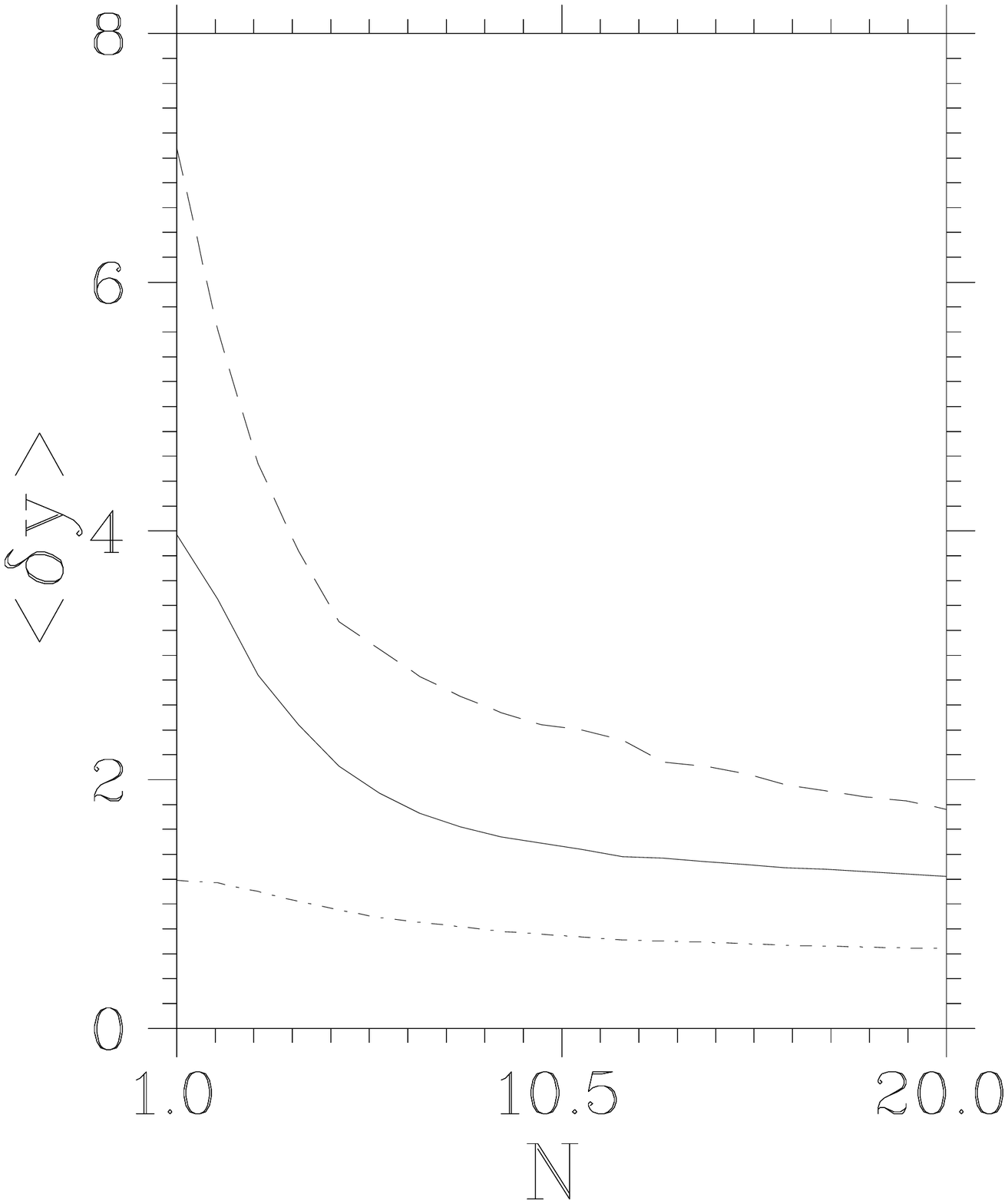}}
\nopagebreak
\begin{center}
{\LARGE {\bf Figure 4} }
\end{center} \vspace*{0.5cm}


\begin{thebibliography}{99}
\bibitem{BFKL}\ E.A. Kuraev, L.N. Lipatov and V.S. Fadin, Sov. Phys.
  JETP {\bf 45} (1977) 199; Ya. Ya. Balitskii and L.N. Lipatov, Sov.
  J.  Nucl. Phys.  {\bf 28} (1978) 822;
\bibitem{Lipatov}
   \L.N. Lipatov, Sov. Phys. JETP   {\bf 63} (1986) 904;
    L.N. Lipatov in {\em Perturbative Quantum Chromodynamics} Ed.
    A.H. Mueller, World Scientific Singapore
\bibitem{Marchesini}\ G. Marchesini and B.R. Webber, Nucl. Phys. {\bf B349}
 (1991) 617
\bibitem{Bartels}\ J.Bartels, H. Lotter and M. Vogt, Phys. Lett. {\bf B373}
 (1996) 215
\bibitem{DelDuca93}\ V. Del Duca, M.E. Peskin and W-K. Tang, Phys.
  Lett.  {\bf B406} (1993) 259
\bibitem{Ryskin}\ M.G. Ryskin, Sov. J.  Nucl. Phys. {\bf 32} (1980)
 133
\bibitem{Simonov}\ Y.A. Simonov, private communication
\bibitem{Hancock}\ R.E. Hancock and D.A. Ross, Nucl. Phys. {\bf B383} (1992)
575
\bibitem{Kwiecinski}\ A.J. Askew, J. Kwiecinski, A.D. Martin and P.J. Sutton,
Phys. Rev. {\bf D47} (1993) 377
\bibitem{Fadin}\ V.S. Fadin and L.N, Lipatov,  DESY Report 96-020 (1996) \bibitem{Mueller}\ A. Mueller, Nucl. Phys. {\bf  B415} (1994) 373;
    N.N.Nikolaev, B.G. Zaharov and V. Zoller, Phys. Lett. {\bf B327}
(1994) 149

\end{thebibliography}
\end{document}